\documentclass[conference]{IEEEtran}
\IEEEoverridecommandlockouts
\usepackage{cite}
\usepackage{amsmath,amssymb,amsfonts}
\usepackage{algorithmic}
\usepackage{graphicx}
\usepackage{textcomp}
\usepackage{comment}
\usepackage{xcolor}
\usepackage{soul}
\usepackage{url}
\usepackage[colorinlistoftodos]{todonotes}
\def\BibTeX{{\rm B\kern-.05em{\sc i\kern-.025em b}\kern-.08em
    T\kern-.1667em\lower.7ex\hbox{E}\kern-.125emX}}
\usepackage{mathtools}
\DeclarePairedDelimiter{\ceil}{\lceil}{\rceil}

\newcommand{\inhabitants}{i}
\newcommand{\bl}{b}
\newcommand{\vod}{v}
\newcommand{\hub}{hub}
\newcommand{\corenode}{CN}
\newcommand{\corerouter}{crm}
\newcommand{\edgenode}{EN}
\newcommand{\edgerouter}{erm}

\usepackage{color}
\definecolor{darkgreen}{rgb}{0.01,0.5,0.01}
\newcommand{\gael}[1]{}
\newcommand{\ab}[1]{\textcolor{green}{\textbf{Aurelie:} {\slshape #1}}}

\newcommand{\etal}{\textit{et al.}}
\newcommand{\dlstyle}[1]{\textit{#1}}

\begin{document}

\title{
Assessing VoD pressure on\\network power consumption\\

\thanks{This study has been financially supported by the French Research Agency through the PostProdLEAP project (ANR-19-CE23-0027-01)}
}

 \author{\IEEEauthorblockN{Gaël Guennebaud}
 \IEEEauthorblockA{
 \textit{ Inria, Univ. Bordeaux}\\
Talence, France\\
 gael.guennebaud@inria.fr}
 \and
 \IEEEauthorblockN{Aurélie Bugeau}
 \IEEEauthorblockA{
 \textit{Univ. Bordeaux, CNRS, LaBRI UMR 5800}\\
 \textit{IUF}\\
Talence, France\\
 aurelie.bugeau@u-bordeaux.fr}
 \and
  \IEEEauthorblockN{Antoine Dudouit}
 \IEEEauthorblockA{
 \textit{Univ. Bordeaux, LaBRI, UMR 5800, Inria}\\
Talence, France}
}
\maketitle

\begin{abstract}
Assessing the energy consumption or carbon footprint of data distribution of video streaming services is usually carried out through energy or carbon intensity figures (in Wh or gCO2e per GB). In this paper, we first review the reasons why such approaches are likely to lead to misunderstandings and potentially to erroneous conclusions. To overcome those shortcomings, we propose a new methodology whose key idea is to consider a video streaming usage at the whole scale of a territory, and evaluate the impact of this usage on the network infrastructure. At the core of our methodology is a parametric model of a simplified network and Content Delivery Network (CDN) infrastructure, which is automatically scaled according to peak usage needs. This allows us to compare the power consumption of this infrastructure under different scenarios, ranging from a sober baseline to a generalized use of high bitrate videos.
Our  results  show  that  classical  efficiency  indicators  do not reflect the power consumption increase of more intensive Internet usage, and might even lead to misleading conclusions.
\end{abstract}


\section{Introduction}
Internet traffic has grown exponentially in the last decade. Cisco~\cite{Cisco2017} projected an increase of IP traffic from 122 exabytes per month in 2017 to 365 in 2022.  
Studying energy consumption and impacts on climate change of data transmission over the Internet has therefore received much attention over the last decade. 
Such work usually strives to estimate the overall yearly energy consumption of the Internet (in TWh/year) from which energy intensity estimates of data transmission (in Wh/GB) are extracted. A recent review~\cite{network_td_model} of these works shows high variability in the results and even inconsistencies between overall energy consumption, energy intensity, and Internet traffic.
Those variations are likely explained by differences in Internet modeling, system boundaries, hypotheses, and methodologies.


\paragraph*{Limits of energy intensity estimates}
Despite those high uncertainties, such energy intensity estimates are frequently used to assess the energy consumption, or environmental impacts, of data transmission of a given Internet service such as, for instance, video-streaming.
A commonly raised question is: ``what is the energy consumption of transferring one GB of data?'', which exhibits severe limitations.

Firstly, we argue such a question is ill posed as it depends on numerous variables that go way beyond distinguishing the core, fixed-access, and mobile-access networks. Some other variable examples include the technological maturity of the considered network (from old energy-intensive equipment, to the newest generation) or the actual route taken by the data: a data intensive service hosted in the US but used in Europe will have a very different impact on the network than another service hosted in the same city than its primary users.

Secondly, unlike what their units convey, those numbers exhibit a poor proportionality with the physical reality. Indeed, one can quickly come to the false conclusion that, e.g., reducing by two the amount of data of a given usage will reduce by two its impacts. This limitation might not be a problem when those numbers are exclusively used as an \textit{attributional} key to allocate the overall shared footprint across the different usages retrospectively. Their use in a \textit{consequential} manner is, however, very frequent and misleading both for a short or long term point of view~\cite{schien2023help}. On the shorter term because the infrastructure is permanently switched on, and the volume of data passing through it at a given time has very little influence on the power consumption of the equipment (especially for fixed network equipment). On the longer term, one could expect a correlation because if the traffic volume increases, the traffic peak is expected to increase too, yielding to an increase in the infrastructure equipment, and thus an increase of the overall consumption of the infrastructure~\cite{schien2022rethinking}. Conversely, if the traffic is maintained or decreased, oldest equipments might be replaced by smaller and more efficient ones when renewed. However, this long term correlation is only partial because i) only a subset of network hardware is subject to such correlation with peak demands, ii) the energy efficiency of such equipment improves quickly over time, and iii) two identical volumes of transferred data might have very distinct effects on local traffic peaks (because of different bitrates, different routes, or different \textit{burstiness}~\cite{Guerin1991}).

Thirdly, such intensity numbers (in Wh/GB or gCO2e/GB) are only \textit{efficiency} indicators hiding the true \textit{absolute} energy consumption or absolute impacts, which are the only numbers that really matter. This observation combined with the aforementioned second point yield a paradox: increasing the total amount of traffic increases load percentage and enables scaling gains. Both lead to an improvement (i.e, a decrease) of those efficiency indicators, while the absolute impacts increase. In contrast, sobriety behaviors are certain to maintain or decrease absolute impacts even though they might degrade those efficiency indicators.
%

Fourthly, such indicators tend to put the responsibility of the impacts solely on the consumer side, while we believe environmental impacts are systemic problems that concern manufacturers, content providers and users. In other words, by hiding the global absolute impacts, such indicators tend to lead to individualization at the expense of a collective vision at which different and more effective levers could emerge.


\gael{add "tricky allocation issues of shared or multi-purpose equipment"? c'est déjà discuté dans Related works, et l'intro est déjà très longue, donc potentiellment suffisant...}

\paragraph*{Contributions}
To address those shortcomings and convey a more realistic understanding of data transmission, we propose a new methodology whose central idea is to consider a given Internet usage at the whole scale of an appropriately chosen territory, and evaluate the impact of this usage, or variants of this usage, on the IT infrastructure. This is accomplished through a parametric bottom-up network modeling of a simplified network infrastructure. Starting from a minimalistic infrastructure (the \textit{baseline}), our model automatically scales the required hardware according to peak usage scenarios, from which absolute power and energy consumption can be estimated and compared to the baseline or other scenarios.
%
Our simplified model relies on a tree representation of the network infrastructure, allowing us to adjust it to peak access rates in a hierarchical manner. 


Having a precise estimate of the overall energy intensity of the Internet is out of the scope of this paper. The version of the parametric model we propose in this paper is rather designed to analyze and compare given use-cases relatively. It is an approximation of the reality with the aim of comparing past and future scenarios for a known usage under the same boundaries. For the sake of simplicity and clarity, we restrict ourselves to the scale of a \textit{territory} with a limited geographical extent. The general principle of our methodology is presented in Section~\ref{sec:methodo}.


In Section~\ref{sec:vod}, we demonstrate our methodology on video-on-demand (VoD) at the scale of the France territory. This use case implies a high data traffic playing an important role on the scale of current infrastructures. Through our experiments, we propose an evaluation of the impacts that would result if watching only HD video in contrast to higher quality streams. Our model enables analyzing which network equipment is most impacted by each parameter of the scenario. The proposed methodology is therefore a first step towards a better understanding of the consequences of politic, industrial or societal decisions on infrastructure sizing. In particular, our results confirm the aforementioned claims and paradox, and show that using a simple efficiency indicator may lead to misleading decisions. Owing to the lack of data, in this paper we restrict our analysis to the estimation of power consumption, however, our model could easily be extended to account for the manufacturing and other life-cycle phases to estimate other environmental indicators.



\section{Related works}

This section presents prior related works on energy consumption of data transmission. 
The ICT infrastructure is commonly decomposed into three parts: datacenters, transmission networks and user equipment. Many works have tackled the estimation of energy intensity for each part. We here mainly focus on the network. Note that some earliest works estimating the energy consumption of the Internet included datacenters while others did not, but it is now more common to consider them apart~\cite{coroama2015}. Early works~\cite{Koomey2004,Gupta2003} were bottom-up approaches. They were based on an inventory of all US computing and networking equipment and their average yearly energy consumption to compute the total energy of networking devices in the US. Those overall energy consumption were then normalized by some estimates of the overall traffic volume, yielding to energy intensity indicators. Such indicators have been re-evaluated on a regular basis, and we refer to Aslan \etal~\cite{Aslan2018} and Coroama~\cite{network_td_model} for recent surveys and analysis. In the rest of this section, we rather focus on network models and alternative approaches.
%
%

%
Baliga \etal~\cite{baliga_2009} proposed one of the first bottom-up network infrastructure model. It is composed of the access, metro and core networks. The total power is the sum of the power $P_i$ of each equipment multiplied by i) the power usage effectiveness (PUE), ii) a redundancy factor $\eta$ to ensure functionality in case of failure, and iii) a scaling factor based on peak access rate $R_i$ over individual capacity $C_i$:
\begin{align*}
P = \sum_i {PUE \times \eta \times P_i \times \ceil[\bigg]{\frac{R_i}{C_i}}} .
\end{align*}

Baliga's model has inspired many later works~\cite{schien_2015, coroama_2013, Hinton2015, coroama2015, IEA2019}.
In particular, Hinton \etal~\cite{Hinton2015} presented an extension to assess the energy consumption of optical networks for different services and scenarios. Each network element is associated with an affine power profile distinguishing the static power $P_{idle}$ and the linear component which is assumed to be proportional to the current throughput (in bit/s). They observed that for such fixed network elements, this proportional part is very small with $P_{idle} > 0.9 P_{max}$. This observation is also confirmed by Malmodin~\cite{power_model}. With such an affine power model, allocating the proportional part boils down to a simple volume-based allocation. To accommodate for the idle power, they proposed different allocation strategies, and in particular one based on relative throughputs.
Malmodin's power model~\cite{power_model} is also based on an affine power-profile. The idle power is first equally spread to each line, and then distributed to the potentially multiple users and devices using this line, which can be rather sketchy to do in practice.
In a similar vein, Ullrich \etal~\cite{Ullrich2022} described an hybrid allocation: duration-based for the customer premises equipment (CPE) and access network equipment, and volume-based for the core network.
All those strategies~\cite{Hinton2015,power_model,Ullrich2022}, however, ignore the energy consumption during standby time, while the last twos assume that the idle power consumption is unrelated to traffic demand, which tends to artificially minimize the network part of the impacts of a given usage.
Our approach overcomes those shortcomings by replacing the allocation issues by a more systemic view, and estimating the ideal idle-power for a given usage or service.


To better understand network energy consumption or GHG emissions, several studies focused on a narrow and specific use case. For instance, Schien \etal~\cite{schien2012} used traceroute data to estimate the number and type of network devices involved in digital media transmission. Coroama \etal~\cite{coroama_2013} considered a 40 Mbps videoconferencing transmission between Switzerland and Japan, and modeled all Internet nodes and links along the way, distributing the energy according to the relative traffic volumes.
Ficher \etal~\cite{fisher_2021} estimated the carbon footprint of transmitting one gigabyte of data on a specific segment of the RENATER network.
Golard \etal~\cite{Golard2022} evaluated and projected the total energy consumption of broadband radio networks at the scale of Belgium.
Another use-case that has received attention is the assessment of the carbon footprint of watching one hour of video streaming, as discussed in Section~\ref{sec:vod}.

\section{Methodology}
\label{sec:methodo}

In this section, we present a general overview of our methodology. A concrete instance on the VoD streaming use case is given in Section~\ref{sec:vod}.
Let us recall that our goal here is not to estimate the impact of an existing infrastructure, but rather to estimate the impacts of a given service or use-case through its pressure on the dimensioning of a \textit{hypothetical} infrastructure.
It is instanced by a bottom-up model that automatically scales the infrastructure to different \textit{scenarios} and hypotheses. From this infrastructure, we can then estimate its \textit{absolute} power consumption and other environmental impacts.
To estimate the absolute impacts of a given use-case, one starts to define a minimalist \textit{baseline} scenario from which a baseline infrastructure is generated and evaluated. Then, a second infrastructure is generated and evaluated for the given use-case and the difference between the absolute impacts of these two scenarios is attributed to this use-case. In practice, this approach also permits to compare different hypotheses for the same use-case, hence enabling a better understanding of the consequences of different choices on the infrastructure.
Through this exercise, it is important to consider the whole service/use-case at the scale of a large-enough \textit{territory} to be representative of the service/use-case at hand.
Those few methodological principles are key to avoid the pitfalls discussed in the introduction, but also to avoid tricky allocation issues of shared or multi-purpose equipment.



Below we present our methodology as four main steps. 

\subsection*{Step 1 - Use-case}
In order to guide the next steps, we first need to define the service or use-case that we aim to model, evaluate, and analyze. Examples encompass video streaming, videoconferencing, large file downloading, email communication, etc. In addition, one also has to identify the main parameters and variables associated to this use-case, and their range of values that will be explored (e.g., video resolutions, number of viewers, server localization, file sizes, frequencies, etc.).
At this step, one can already define the \textit{baseline scenario} through the choice of the most sober values for those variables (e.g., no streaming, a few emails a day without attachments, etc.).


\subsection*{Step 2 - Boundary}
This step covers two aspects. First, which parts of the Internet infrastructure are included: datacenters, core, edge, fixed and/or radio network, fiber and/or copper, customer premises equipment (CPE), etc. Second, which geographical territory: a city, a country, the world? The choice of a territory might be dictated by the purpose of the evaluation, e.g., one might be interested in evaluating a service for a given country. Otherwise, for the sake of simplicity, it might be wise to choose the smallest possible territory that is representative of the scenarios identified in step 1.

\subsection*{Step 3 - Design of the parametric infrastructure model}
This step consists in designing the parametric model that will generate infrastructures according to some dimensioning variables. To this end one must start to design a minimalist infrastructure, e.g., every home and datacenters of the considered territory must be connected with the capacity to exchange some bits, the radio network must cover 99\% of the territory, every user of our scenarios possesses at least one smartphone, tablet or laptop, etc. This minimalist instance completes the baseline scenario identified at step 1. The model has to be designed to be able to scale-up and to cover the range of use-cases and boundaries defined in the previous steps.

\begin{figure}[b]
\centerline{}
\includegraphics[width=0.45\textwidth]{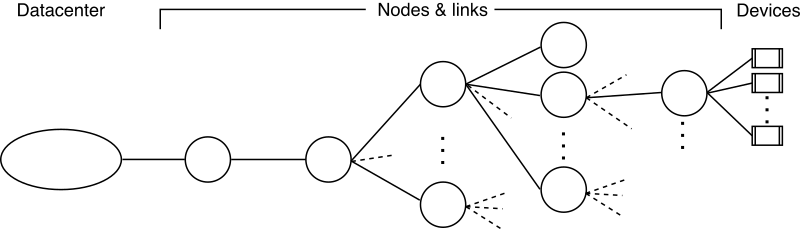}
\caption{Abstract tree representation of the infrastructure. In this example, tree goes from a main datacenter to many devices through nodes and links. A node can, for instance, represent an IXP that includes a CDN or routers. The detailed infrastructure for our use-case is presented in Figure \ref{fig:network}.}
\label{fig:tree}
\end{figure}

\begin{figure*}[t]
\centerline{
 \includegraphics[width=0.75\textwidth]{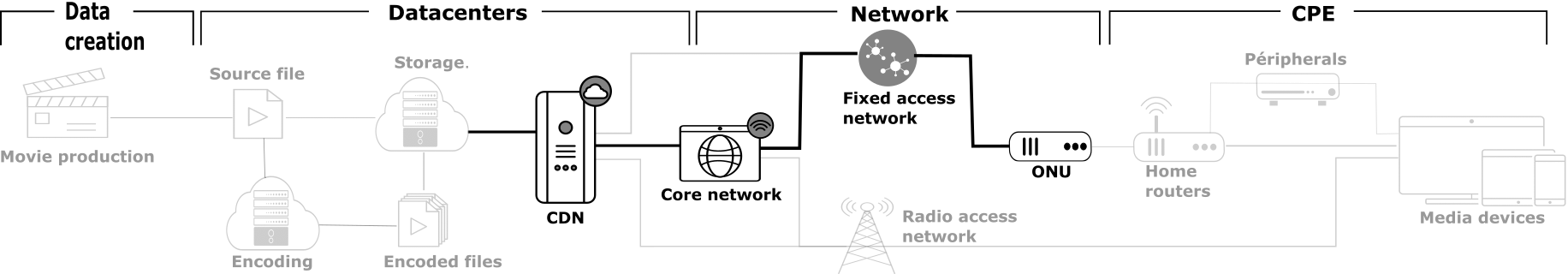}}
\caption{General OTT VoD infrastructure. Steps in black are the one included in our boundary. Figure modified from~\cite{DIMPACT}.}
\label{OTT}
\end{figure*}

For the network, as a proof of concept, in this work we propose to use a simplified tree representation that goes from the user houses up to the main datacenters hosting the considered service, and passing through nodes representing the different pooling points of the fixed-access, edge, and core network layers. More complex structure representations shall be used in future work. Figure~\ref{fig:tree} illustrates part of the tree of the infrastructure.
Putting aside the end-user devices, the leaves correspond to the home-routers, and the nodes represent converging/splitting points where congestion points might occur. Edges represent fiber links which might either include passive-only equipments, but also active equipments for long-distance hops.
Following previous bottom-up network models, its main dimensioning variables are the bandwidth capacities required at the different nodes and links of the tree.
At each node and link $l$, the equipment is scaled up to the minimal quantity enabling:
\begin{itemize}
    \item[i)] a connection to every subscribers (households)
    \item[ii)] a bandwidth capacity equal or greater to $R_l$ (in bit/s) which has to be adjusted with respect to the peak access rate estimated at the node or link $l$ for the given scenario (see Step-4). Throughout this paper, peak rates are assumed to refer to averaged traffic rates over a few seconds.
\end{itemize}
This quantity is then multiplied by the redundancy factor $\eta$.
The global power consumption is estimated as a sum over all equipments and facilities of the infrastructure. To this end, each equipment must be associated with a power consumption profile, i.e., a static (or idle) power (in W) and a power intensity factor (in W/Gbps) for the dynamic power consumption part which is assumed to be proportional to the actual traffic.
Because we scale the infrastructure to peak needs, we are able to understand the physical reality and bottlenecks behind power consumption.

\subsection*{Step 4 - Scenario evaluation and peak demand modeling}
Finally, to evaluate one of our scenarios, we need to translate it to the dimensioning variables exposed by the model defined in the previous step. In our case, this mainly requires to estimate the capacities $R_i$ from peak access rates estimated at every node and link of the tree.
This step usually embeds a growing margin factor ($\alpha$ in the following) allowing to anticipate exceptional traffic peaks, and for future growth provisioning.




Our scenarios are constructed based on global average statistics of different kinds. The first kind are expressed in term of percentage of ``active'' users (e.g., percentage of active subscribers for the baseline, or the percentage of simultaneous VoD watchers).
The second kind are numbers such as the average number of inhabitants per house.
Whereas using such statistic means would be sufficient when considering a large pool of inhabitants, they cannot be used to reflect worst case scenario near the leaves of the tree where some equipments are shared by few dozens to a few hundreds of inhabitants only.
For a better accuracy, we propose to consider their respective distributions, say $d^n$, for a given sub number of inhabitants $n$. To neglect the most unlikely occurrences only, we use the smallest quantity $q$ such that the probability of having an occurrence $x$ greater than $q$ is extremely low, i.e., $d^n(x>q)<\epsilon$. We used $\epsilon=10^{-9}$. 

For the statistics of the first kind, defined by a percentage $s$ of ``active'' inhabitants, the distribution $d_s^n$ corresponds to a hypergeometric distribution. Assuming that $n$ is small compared to the total number of inhabitants, it is well approximated by the simpler binomial distribution, and we define the function $q_s(n)$ as the smallest $q \leq n$ such that $d^n_s(x>q)<\epsilon$, which corresponds to the $1-\epsilon$ percentile, which is itself computed through the inverse of the cumulative distribution function (CDF) of $d_n^s$.
Note that for very large $n$, we have $q_s(n) \approx s \times n$, but $q_s(n)$ can be significantly greater than $s \times n$ otherwise (e.g., $q_{3\%}(64)/64=21\%$).

For our statistics of the second kind, such as the number of inhabitants per home, we first have to define the respective discrete distribution $d_{\inhabitants}(x)$, and consider the sum of $n$ random variables having this discrete distribution. Again, we assume that $n$ is small compared to the total number of inhabitants. The resulting distribution $d^n_{\inhabitants}(x)$ is thus obtained by the convolution of $d_{\inhabitants}$ with itself $n$ times. As before, we then define $q_{\inhabitants}(n)$ as the $1-\epsilon$ percentile of $d^n_{\inhabitants}(x)$.
Since there is no closed-form formula for $d^n_{\inhabitants}$, computing $q_{\inhabitants}(n)$ for many different values of $n$ can be very tedious. Instead, we found that $q_{\inhabitants}(n)$ can be very well approximated by a function of the form:
\begin{equation}
\label{eq:approx_d_i}
q_{\inhabitants}(n) \approx \max( a_\inhabitants n + b_\inhabitants n ^ {c_\inhabitants}, n \bar{d_{\inhabitants}})
\end{equation}
where $\bar{d_{\inhabitants}}$ is the mean of $d_{\inhabitants}(x)$.
The three coefficients $a_\inhabitants$, $b_\inhabitants$, $c_\inhabitants$ are found numerically to interpolate three points taken at $n \in \{16, 128, 1024\}$. Values of $d_{\inhabitants}$ for our use-case presented in next section
can be found in Table \ref{tab:distribHabitants}.
\begin{table}[htbp]
    \centering
    \caption{Distribution of inhabitants per house in France in 2019~\cite{insee2019}.}\label{tab:distribHabitants}\vspace{-2mm}
\begin{tabular}{c|cccccc}
     x & 1 & 2 & 3 & 4 & 5 & $\ge$6 \\
     \hline
     $d_{\inhabitants}(x)$ & 0.369 & 0.326 & 0.135 & 0.113 & 0.041 & 0.016 %
\end{tabular}
\vspace{-2mm}
\end{table}

\gael{TODO add plot of the approximation}

\section{Case study: VoD streaming}
\label{sec:vod}

To illustrate our methodology, we consider VoD streaming.
This use-case has already been addressed by many studies~\cite{shift_project,IEA2020,locat_2021,sfu_2021, DIMPACT, Obringer_2021, Ullrich2022, preist2019evaluating}.
All of them, however, focus on estimating the electricity intensity of one hour of video or yearly electricity usage of a video service, while allocating the network part based on volume of data (Wh/GB), or a mix of volume and duration. As motivated in the introduction, those \textit{a posteriori} attributional allocation strategies can hardy be used to predict the impacts of intensifying VoD streaming, nor to understand the real pressure of VoD streaming on the network infrastructure. Applying our methodology to this use-case will thus reveal new insights.

This section follows the four steps presented in the previous section. For the sake of clarity, the steps 3 and 4 are detailed per kind of equipment.
Owing to the lack of power profile data, and since we consider a fixed network, we will focus on the estimation of the static power only, leaving the dynamic part for discussions in Section~\ref{sec:res}.

\begin{figure*}[t]
\centerline{
 \includegraphics[width=0.9\textwidth]{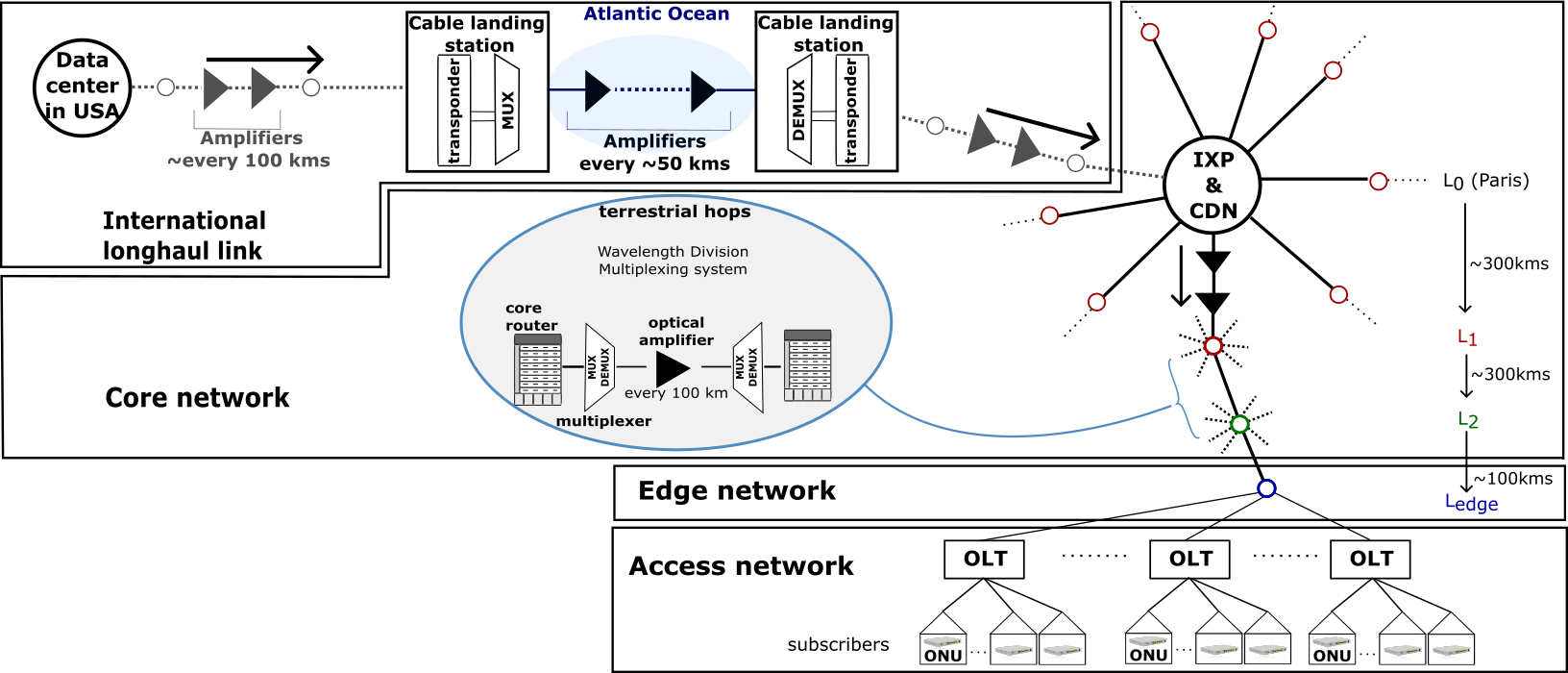}}
\caption{Network modeling of our video on demand use case. We restrict our boundary to fiber to the home network, a unique datacenter and one CDN.}
\label{fig:network}
\end{figure*}

\subsection{Design overview}

\subsubsection{Step 1 - Use-cases definition}
Video streaming in general covers a large range of different types of services, each having their own infrastructure design with different usage patterns. In this study, we limit ourselves to over-the-top (OTT) streaming from a unique service provider for the whole territory. We further assume a bounded catalog controlled by the service provider. We thus exclude other types of video streaming such as Youtube (unbounded catalog), live streaming, IPTV, and advertising videos.

Figure \ref{OTT} depicts the main components of an OTT VoD service. After the content creation stage, the service provider manages its catalog with many different encoding settings and redundant storage in its main datacenters.
In general, videos are not delivered directly from the main datacenters but from closer servers belonging to content delivery networks (CDNs). CDNs are storage servers usually located at Internet service provider (ISP) or Internet exchange points (IXP). CDNs reduce data traffic in the core network (and in particular in submarine and longhaul optical fibers) and improve user experience with faster loads.
Those CDN servers are partly updated every days during low traffic periods.
The videos are then delivered to the customers through the core and access networks.

The main variables are the video quality $R_{\vod}$ (ranging from $0$ to $30$ Mbps), and the percentage $s_{\vod}$ of inhabitants accessing the service simultaneously during the peak period.
Other variables include: the size of the catalog and the percentage of video content accessed through the CDN.

In our baseline scenario we consider a traffic expected to represent a ``minimalist and sober'' use of the Internet as communication and information sharing means, excluding all traffic-intensive usages such as video streams, large file transfers, heavy web pages, etc. (both $R_{\vod}$ and $s_{\vod}$ are set to $0$). We further ignore all the B2B traffic.
At peak hours, this baseline traffic is modeled as a global percentage of active customers $s_\bl=2\%$, and a per customer speed rate $R_\bl=10$ Mbps. Those numbers yields an average peak rate of $200$ kbps per customers, which matches average peak rates for copper lines in 2013 in France. Those numbers are thus rather high for a ``minimalist and sober'' use of Internet, but they can be considered as conservative.
We further assume that the baseline and VoD traffic peaks are fully correlated, which is also a rather conservative choice.



\subsubsection{Step 2 - VoD streaming boundaries}
For the sake of simplicity, we chose a narrow boundary in this study. Parts in dark gray in Figure~\ref{OTT} are those that we include, while we ignore steps in light gray (content creation, encoding, customer management, end-user devices, ...).
%
To be consistent with our tree structure, we consider a unique CDN that includes both servers and dedicated routers.

\noindent\textit{Territory:}
The territory is chosen to be representative of the geographical scale typically covered by a unique CDN.
This makes Metropolitan France, which hosts the ICT4S 2023 conference, a rather good candidate to scale our network scenarios. We thus consider $65e6$ inhabitants for about $\#home=30e6$ households,
each having an internet connection and being a customer of the VoD service. We assume that 2/3 of our baseline traffic stays in this territory. The CDN is naturally located in an IXP in Paris.



\noindent\textit{Main datacenters:} The main datacenter servers and routers of the VoD service provider are expected to be shared by many countries. We thus chose to ignore their own power consumption. However, in order to account for the Internet traffic load required to update our CDN, as well as to deliver videos that are not cached in the CDN servers, we still consider one main datacenter located in North America at about 900km from the Atlantic submarine cable landing point. We also assume that the 1/3 of the baseline traffic coming from international sources goes through this same route.


\noindent\textit{Network:} We include both the core and edge network active equipment, but ignore passive ones. For the access network, since the VoD service we consider is mostly used at home, we consider only a fixed-access network that we assume to be fully implemented through the GPON (Gigabit Passive Optical Network) FTTH (fiber-to-the-home) architecture.

\noindent\textit{CPE:} Since our study focuses on the network, on the customer side, we consider only the ONU needed for the GPON fiber architecture, but ignore all other devices such as home-routers, set-top-boxes, TVs, laptops, etc.


\subsubsection*{Summary of the infrastructure}
Figure~\ref{fig:network} summarizes the infrastructure for our use-case. Our infrastructure and the number of equipments, is scaled according to peak usage that depends on the scenario. In the next subsection, we detail each node and how we compute the quantity and power of all equipments.  Some general parameters are considered. The $PUE$ indicator for the network is set to 1.8~\cite{Aslan2018} for all experiments, while the redundancy factor in case of failure is $\eta=2$.
Our design is largely inspired by the model of Baliga et al.~\cite{baliga_2009}, some power consumption coming from this paper. When it is the case, we update energy intensity (W/Gbps) of equipments considering the energy efficiency gains through years. The formula proposed by the authors: $ I = \frac{P}{C} = \frac{P_0}{C_0} \times(1-\gamma)^t$ considers the $t$ years between 2020 (our reference year) and 2008. $P_0$ is power and $C_0$ capacity from 2008 and $\gamma=0.1$.

\begin{figure}[t]
\centerline{
 \includegraphics[width=0.9\linewidth]{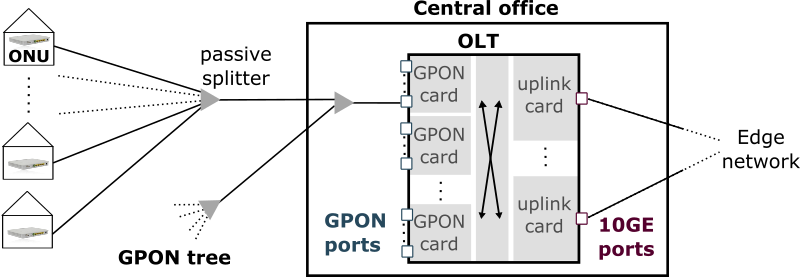}}
\caption{Representation of the fiber to the home structure that connects core network to access network with GE and GPON ports.}
\label{fig:gpon}
\end{figure}

\subsection{GPON based FTTH accesss network} 

\subsubsection*{Step 3 - Access network design}
To model FTTH, we have chosen the GPON architecture as our outermost link (Figure~\ref{fig:gpon}).
Its main component is the Optical Line Terminal (OLT) hosted in central offices (hubs) of the operator. It is primarily composed of GPON cards with up to $\#gpon/card = 16$ GPON ports per card, and up to 16 cards per OLT. Each GPON port has a maximal capacity of $2.5$Gbps shared by $\#sub/gpon \leq 128$ subscribers through a unique output fiber. This fiber is then split in a tree structure through passive optical elements. This tree ends with one Optical Network Unit (ONU) in each household. This active equipment is required to deal with the temporal multiplexing of the underlying GPON protocol.
In our model, the OLT is connected to the backbone through 10 Gigabit Ethernet ports (10GE ports). We have at least one 10GE port per OLT and thus at most $256\times \#sub/gpon$ subscribers per 10GE port.
The number $\#\hub$ of central offices is fixed and determined to cover the whole territory.
An opendata list of central offices in France in 2020\footnote{\url{https://www.data.gouv.fr/fr/datasets/localisations-des-noeuds-de-raccordement-abonnes-nra-et-optiques-nro-dans-openstreetmap/}} reports about 1500 hubs dedicated to OLT hosting, and 4300 other ones tagged as ``copper''. In practice, many of them also host OLTs, and the presence of four main operators also tends to increase the number of hubs compared to a single operator scenario. All in all, we took $\#\hub=3000$ (in practice this parameter has a very small influence on the overall results).
For the sake of simplicity, we also assume that the subscribers are equally spread among the hubs.
Static power consumption and capacity of those equipments are given in Table~\ref{tab:powerdata}.
\begin{table}[b]
    \centering
    \caption{Static power $P$ and capacity $C$ of some network equipment. Numbers are adapted from~\cite{baliga_2009} and~\cite{OpenConnect}.}\label{tab:powerdata}
\begin{tabular}{c|cc}
  & Power (W) & Capacity \\
  \hline
  ONU & $P_{ONU} = 2.5$ & - \\
  GPON port & $P_{GPON} = 15$ & $C_{GPON} = 2.5$ Gbps \\
  10GE port & $P_{GE} = 30$ & $C_{GE}   =  10$ Gbps \\
  Ethernet switch module & $P_{eth} = 60$ & $C_{eth}   =  40$ Gbps\\
  BNG module & $P_{BNG} = 75$ & $C_{BNG}  =  40$ Gbps\\
  Edge router module & $P_{\edgerouter} = 120$ & $C_{\edgerouter}  =  40$ Gbps\\
  Core router module & $P_{\corerouter} = 1400$ & $C_{\corerouter}  = 560$ Gbps \\
  Flash server & $P_{flash} = 320$ & $C_{flash} = 190$ Gbps \\
  Storage server & $P_{sto} = 400$ & - 
\end{tabular}
\end{table}

From this design, the total number of GPON ports is:
\begin{equation}
\#gpon = \left\lceil \frac{\#home}{\#sub/gpon} + \#\hub \times \#gpon/card \right\rceil .
\end{equation}
The right-hand side term accounts for the fact that GPON cards are, in practice, not completely filled. We therefore add the equivalent of one GPON card with $\#gpon/card=16$ ports per hub.
%
The number of 10GE ports is scaled with respect to the peak-rate demand $R^*_{olt}$ at one OLT:
\begin{align}
\#{ge} = \eta \, \#{olt} \left\lceil \frac{R^*_{olt}}{C_{GE}} \right\rceil \,.
\end{align}
The number of OLTs is obtained as:
\begin{align}
\#{olt} = \max\left( \#{hub}, 
\left\lceil \frac{\#{gpon}}{\#{gpon/card} \times \#{card/olt} } \right\rceil
\right)  \,,
\end{align}
with $\#{card/olt}$ the maximal number of GPON cards per OLT ($\#{card/olt}=16$).
%
The total power of the access network is the cumulative power of all GPON and GE cards:
\begin{align}
P_{access} =  PUE \times \left( \#gpon \times P_{gpon}
+ \#ge \times P_{ge}  \right).\label{eq:Paccess}
\end{align}



\subsubsection*{Step 4 - Access network peak demand}
In order to evaluate the above equations, we have to compute the actual maximal number of subscribers per GPON tree ($\#sub/gpon$) such that the peak demand traffic of a given scenario does not exceed its bandwidth capacity, as well as the demand rate $R^*_{olt}$.
To this end, we will exploit the statistical tools presented in Section~\ref{sec:methodo}-Step-4.
%
%
%
Since the GPON technology cannot support more than 128 subscribers per GPON port, $\#{sub/gpon}$ is the largest integer $n \leq 128$ such that:
\begin{align}
\alpha_t R(n) < C_{GPON} \,,
\label{eq:gpon}
\end{align}
where $R(n)$ is the peak demand for a pool of $n$ subscribers:
\begin{align}
R(n) = R_{\vod} \times \frac{q_{s_{\vod}} \circ q_{\inhabitants}(n)}{S} + R_{\bl} \times q_{s_{\bl}}(n) \,.
\label{eq:peakrate}
\end{align}
We recall that $\alpha_t$ is a growing margin factor (Section~\ref{sec:methodo}-Step-3), $s_{\bl}$ is the average percentage of active subscribers with baseline usage, $s_{\vod}$ is the average percentage of VoD viewers among the population at peak hours, and $S$ is the average number of viewers sharing the same device/video flux ($S=1.5$~\cite{DIMPACT}). We used $\alpha_t=1.5$.
In practice, we compute this solution using a binary search while accelerating the evaluations of $q_{s_\vod}\circ q_{\inhabitants}(n)$ and $q_{s_0}(n)$ by approximating both of them by functions of the same form as equation~\eqref{eq:approx_d_i} with coefficients computed using the same three points interpolation strategy.
%
The logic for the 10GE uplink ports is slightly different. Indeed, once the number of subscribers per GPON tree is known, the number of subscribers per OLT is fxed, and we thus avoid the need for the binary search by directly setting the peak-rate demand:
\begin{align}
R^*_{olt} = \alpha_t R\left(\left\lceil {\#{sub}} / {\#{olt}} \right\rceil\right) \,.
\end{align}

\subsection{National edge and core network}

This subsection focuses on the part of the network tree connecting the OLTs to the main IXP located in Paris.
The link from the IXP to North America will be addressed in the next subsection.

\subsubsection*{Step 3 - Edge/core network design}
To connect the entire country to the main IXP, we use a tree topology that interconnects children to core router nodes with a star topology. We consider that each core router node has 8 child nodes, and 3 core levels $L_0$ to $L_2$, with $L_0$ corresponding to the main IXP in Paris.
A core router is made of a variable number of modules. Given a capacity requirement of $R$ Gbps, the actual number of modules is given by:
$$\#\corenode(R) = \left\lceil \frac{R}{C_{\corerouter}} \right\rceil,$$
with an electrical power of:
$$P_{\corenode}(R) = PUE \times \eta \times \#\corenode(R) \times P_{\corerouter}.$$
Each of the 64 core nodes of level 2 are then linked to 8 edge nodes (level 3).
An edge node is composed of a modular edge router, a modular broadband network gateway (BNG), and a modular Ethernet switch connected to the OLTs.
The actual quantity of modules composing one edge node and its respective power is computed just as $\#\corenode(R)$ and $P_{\corenode}$ but using the capacity and power features given in Table~\ref{tab:powerdata}:
\begin{align}
P_{\edgenode}(R) = PUE \times \eta \times \sum_{k \in \{eth,BNG,\edgerouter\}}
  \left\lceil \frac{R}{C_{k}} \right\rceil \times P_{k}.
\nonumber
\end{align}


Core and edge routers are connected through a wavelength division multiplexing (WDM) transport system, composed of two terminal multiplexers (one at each extremity, with a power of 4.6 W per channel), amplifiers every 100 km (with a power of 3.5 W per channel), and a capacity of 40Gbps per channel.
For a capacity $R$ and distance $dist$, the electrical power of such a link is thus:
\begin{align*}
\small
  P_{wdm}(R, dist) = \eta \ceil[\bigg]{{\tiny\frac{R}{40}}} \Bigg( 
    PUE\times 9.2 
  + \ceil[\bigg]{{\tiny\frac{dist}{100}} - 1}\times 3.5\Bigg).
\end{align*}

Let $R_l^*$ be the required capacity at a node level $l$, and $dist_l$ be the average distance separating a pair of nodes at level $l-1$ and $l$. In our design, $dist_1=dist_2=300$ km, and $dist_3=100$.
The overall power $P_{nat}$ of national core and edge nodes and links are finally:
\begin{align*}
P_{nat} &= 2^3 \times {P_{\edgenode}(R_3^*)}\\
&+ \sum_{l\in [0,2]} 2^l \left( {P_{\corenode}(R_l^*)} + {P_{wdm}(R_{l+1}^*, dist_{l+1})} \right) .
\end{align*}



\subsubsection*{Step 4 - Edge/core network peak demand modeling}
The capacity $R_l^*$ required at each level $l$ is estimated from the peak demand of our scenarios in a similar fashion than for the uplink ports of the access network.
Observing that the number of subscribers related to one node is equal to $\#home / 2^l$, we directly have: $R_l^* = \alpha_t R\left( \#home / 2^l \right)$.
\subsection{International longhaul link}

\subsubsection*{Step 3 - International longhaul design}
The main IXP is connected to the main datacenter in North America through a longhaul WDM transport system made of two terrestrial parts of 600 and 900 km respectively, and one WDM submarine section of 8000km.
We further assume that about 7 core nodes are crossed over along this route. 
Let $R_u^*$ be the required capacity for this line.
The electrical power for the terrestrial part is thus:
\begin{align*}
P_{intT} = 7 \times {P_{\corenode}(R_u^*)} + \sum_{dist\in \{600,900\}} \bigg( P_{wdm}(R_u^*, dist) \bigg). 
\end{align*}

Submarine WDM systems are slightly more complex to model. It is composed of two terminal multiplexers (35W per channel), with repeaters (0.2W per channel) placed to amplify the signal every 50 kilometers and a capacity of 40 Gbps per channel. The repeaters are powered by electrical suppliers located on the coast with 80\% energy efficiency through cables having a resistance yielding power loss of about $0.004$ W/km~\cite{baliga_2009}. This sums up to $142$ W per channel for our 8000 km cable, yielding for the final electrical intensity of the undersea connection:
\begin{align}
P_{intU} = PUE \times \eta \times \alpha_u \times \ceil[\bigg]{\frac{R_{u}^*}{40}} \times 142.
 \label{eq:Uf}
\end{align}
Submarine cables are often designed with a larger margin factor than terrestrial links, hence the dedicated scale factor $\alpha_u=2$~\cite{baliga_2009}.

\subsubsection*{Step 4 - International longhaul peak demand modeling}

The peak demand rate $R_{u}^*$ between the main datacenter and our main IXP/CDN is the maximum between the peak rate $R_{fill}^*$ to fill CDN servers and the peak traffic rate corresponding to the baseline and VoD scenarios.
For the later, we assume that only $1/3$ of the baseline traffic goes trough this international link, and that $\%CDN$ percent of the VoD traffic is handled by the CDN (we used $\%CDN=80\%$).
We end up with:
\begin{align}
R_{u}^* = \max\left(R_{fill}^*, (1-\%CDN) R_{\vod}^*\right)
 + \frac{1}{3}  R_{\bl}^* ~.
\label{eq:Rcore}
\end{align}
This equation assumes that the hours of CDN filling might overlap with the baseline peak, which is unlikely but conservative.
The quantities $R_{\bl}^*$ and $R_{\vod}^*$ are the global peak rates for all customers for the baseline and VoD usages respectively:
\begin{align*}
  R_{\bl}^* = s_{\bl} \times \#home \times R_{\bl} ~;
  R_{\vod}^*  = s_{\vod} \times \bar{d_{\inhabitants}} \times \#home \times R_{\vod} ~.
\end{align*}

\subsection{CDN}

\subsubsection*{Step 3 - CDN design}

Our CDN includes different kind of servers. Following Netflix's CDN design~\cite{OpenConnect}, we consider a CDN made of a few \textit{storage} servers having a large storage capacity to hold a large percentage of the catalog content, and many \textit{flash} servers having a limited storage capacity, but a very high throughput.
It is completed with dedicated edge routers, yielding an overall electrical power modeled as:
\begin{align}
P_{CDN} = PUE \times \bigg(& \ceil[\bigg]{\frac{R^*_{CDN}}{C_{flash}}} P_{flash} + \#sto ~ P_{sto} \nonumber \\
 & + \eta \ceil[\bigg]{\frac{R^*_{CDN}}{C_{\edgerouter}}} P_{\edgerouter} \bigg) ~.
\label{eq:Pcdn}
\end{align}
where $R^*_{CDN}$ is the required throughput capacity.
The average power and throughput of the flash and storage servers are given in Table~\ref{tab:powerdata}.


\subsubsection*{Step 4 - CDN peak demand modeling}
Following Netflix documentation~\cite{OpenConnect}, we set the number of storage servers to $\#sto=40$, and kept it fixed for all our VoD scenarios even though one could slightly adjust it with respect to the maximal video bitrate.
The throughput capacity $R^*_{CDN}$ is estimated from the global peak demand $R_{\vod}^*$, and the percentage $\%CDN$ of content effectively provided by the CDN:
\begin{align*}
  R_{CDN}^* = \alpha_t \times \%CDN \times R_{\vod}^* \,.
\end{align*}
Finally, we found that the bitrate $R^*_{fill}$ needed to update the $\#sto$ CDN storage servers on a daily basis to be quite negligible. According to Netflix~\cite{OpenConnect}, $\sim{}1.8\%$ of the content of CDN storage servers are updated every day. With 320 TB of capacity each, and assuming they are updated during a period of 8h, this yields $R_{fill}=72$ Gbps, which is $\sim{}30$ times lower than our estimated baseline peak for the international link.

\section{Results on different scenarios}
\label{sec:result}
In this section, we compare our baseline with other scenarios. We first analyze the influence of video quality and discuss proportionality (Section~\ref{sec:res}), and then evaluate an hybrid OTT+DTT scenario (Section~\ref{sec:locat}).
We implemented our model as a static web-application allowing the user to modify all the parameters and hypotheses of our model, including the capacity and static power of each network element. This application will be made available online upon acceptance.

\subsection{Effect of video quality}
\label{sec:res}

With our consequential methodology, we wish to understand what could be or have been the consequences of some restrictions over infrastructure dimension and energy consumption. In this section, we propose to compare the influence of several video qualities,
namely HD (1280$\times$720 at 3 Mbps),
FHD (1920$\times$1080 at 5 Mbps),
UHD (3840$\times$2160 at 16 Mbps),
and a fourth UHD++ scenario with 4K resolution, high-dynamic-range (HDR) color depth, and 60 frame-per-second (at 27 Mbps).
In all our scenarios, we assume a baseline peak rate $R_{\bl}=10$ Mbps for a percentage $s_\bl=2\%$ of subscriptions.
All our VoD scenarios assume a peak percentage $s_\vod=20\%$ of viewers through OTT. We recall that contrary to $s_\bl$, this percentage is relative to the whole population.


Moreover, in order to understand the effect of different usage patterns, we also sketched a fictitious download scenario (``DL'') of very large files such as OS updates and AAA video games. The latters become larger and larger with the top 26 ranging from 64GB to 200GB at the end of 2021~\cite{videogametop26}, while generating heavy loads on the network at every release or patch update of the most famous titles.
Our VoD architecture can easily be adapted to such a use-case by setting: $R_{\vod}=200 Mbps$, $S_\vod=3\%$, $\%CDN=95\%$, and replacing the VoD term of equation~(\ref{eq:gpon}) by a simpler subscriber-based term: $R_{\vod} q_{s_{\vod}}(n)$. For this term, we also relaxed the confidence parameter to $\epsilon{}={}10^{-7}$ (Section~\ref{sec:methodo}-step-4), hence accepting that a few users will very likely experience slightly degraded download rates.

Applying the model detailed in the previous section to our scenarios, and integrating static powers over one year, leads to annual power consumptions presented in Table~\ref{tab:power}. This table also reports the percentage of energy consumption increase relative to the baseline scenario.
\setlength{\tabcolsep}{3pt}
\begin{table}[b]
    \centering
    \vspace{-4mm}
    \caption{Annual power consumption in $GWh$ for each network part.}
    \label{tab:power}
    \begin{tabular}{c|c|cccc|c}
     Scenario& ONU &  Access & National  & Int.     & CDN & Total  \\
             &     &         & Core+Edge & longhaul &     &  \\
    \hline
    Baseline &  667 &  69 &  9     & 2.8   &   0   &  748       \\
    HD       &  667 &  71 &  17    & 10    &  2.2  &  767 (+3\%)\\
    FHD      &  667 &  72 &  25    & 15    &  2.5  &  782 (+5\%)\\
    UHD      &  667 &  84 &  70    & 43    &   11  &  875 (+17\%)\\
    UHD++    &  667 & 136 & 114    & 70    &   18  & 1005 (+34\%)\\
    \dlstyle{DL}  &  \dlstyle{667} & \dlstyle{271} & \dlstyle{140} & \dlstyle{2.8}  &  \dlstyle{32} & \dlstyle{1113 (+49\%)}
    \end{tabular}
\end{table}

A first observation is that the overall energy consumption is not directly proportional to bitrates. This is especially true when including the ONUs that are plugged 24/7 in every home.
The increase of the access network is negligible for HD/FHD bitrates, and it remains limited even for the FHD scenario.
This is because the baseline $1{:}128$ configuration of the GPON trees is enough to handle such bitrates. Only the 10GE uplinks has to be upscaled. The UHD++ scenario, however, yields a much higher pressure on the GPON trees that have to be upscaled to a $1{:}75$ configuration.
For all scenarios, we observed that the submarine cable counts for about 33\% of the whole international longhaul connection.

\begin{figure}[tbp]
\centerline{}
\includegraphics[width=\linewidth]{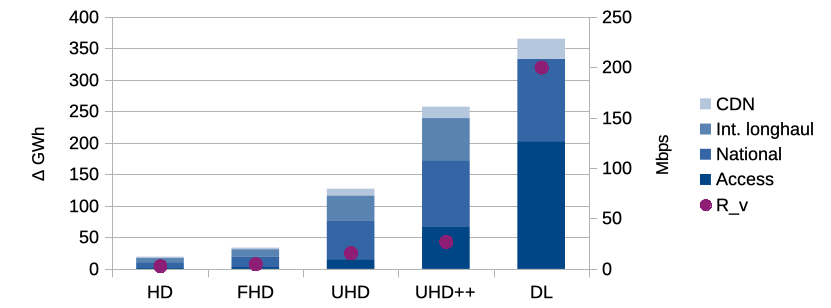}
\vspace{-5mm}
\caption{Energy consumption difference between our VoD/DL scenarios and the baseline. The respective video bitrates $R_{\vod}$ are shown as dots. \label{fig:diff} \vspace{-5mm}}
\end{figure}

Figure~\ref{fig:diff} shows the energy consumption difference between our VoD/DL scenarios and the baseline.
One can observe that for the VoD scenarios, for which only the streaming bitrate changes, this difference is much more proportional to the VoD bitrate than absolute consumption, though not perfect. However, when including a different use-case as our large-file download scenario, this apparent correlation breaks.



Owing to the lack of precise power consumption profiles for each of the considered equipments, the previous energy estimates cover static power consumption only, ignoring the dynamic part. In order to get a rough estimate of what could be the effect of accounting for the dynamic power consumption, we used the average dynamic intensity factor estimated by Malmodin~\cite{power_model} for a fixed line (i.e., $\sim{}0.1$ Wh/B). Considering an average of 2GB (resp. 25GB) of data per subscriber per month for the baseline (resp. DL) use-case, and an average of 3.2h of OTT video per day per subscriber, we obtained the total yearly traffic volume and updated yearly energy consumptions reported in Table~\ref{tab:efficiency}.
As expected, for fixed lines the dynamic power consumption part is quite negligible.
Comparing the UHD++ and DL scenarios, we see that both the absolute (1048 vs 1114 GWh) and incremental (300 vs 366 GWh) energy consumptions are even less correlated to volume (433 vs 10 EB) than bitrates (27 vs 200 Mbps).
This table also reports the energy intensity indicators obtained as the ratio of those two numbers in Wh/GB.
This clearly shows that increasing the video bitrate enables decreasing the relative data transmission consumption when expressed in Wh/GB, confirming that such an efficiency indicator does not reflect the power consumption increase of more intensive Internet usage.


\subsection{Evaluation of DTT caching}\label{sec:locat}

The LoCaT report~\cite{locat_2021} estimates GHG emissions associated with serving TV content across different platforms, including digital terrestrial television (DTT), IPTV and OTT. Their functional unit was delivering one hour of video to a TV set.
In one of their scenario, the authors studied the opportunity of viewing VoD content through DTT broadcasting. This requires one additional home-caching device to every home, which would enable receiving content via the DTT signal, cataloging, and storing content locally. When the user requests a given video, the device would first search for the content locally and would only stream it through OTT if absent. 
In a sense, this home-caching device plays the role of a personal CDN filled through DTT.
The authors forecast a reduction of about 25\% of OTT downloads, and devices exhibiting a stand-by power of 0.5W, and on power of 10W during 3.5h per day. This yields an average power of 1.9W.

\begin{table}[t]
\centering
\caption{Efficiency estimation based on yearly volume\\ and energy consumption.}
\label{tab:efficiency}
\begin{tabular}{c|ccc}
Scenario & Yearly volume (EB) & Energy (GWh) &	Wh/GB \\
\hline
Baseline    & 0.73	& 748	& 1024 \\
HD          & 49	& 772	& 15.8 \\
FHD	        & 81	& 790	& 9.7 \\
UHD	        & 257	& 901	& 3.5 \\
UHD++	    & 433	& 1048	& 2.4 \\
\dlstyle{DL}          & \dlstyle{10}	& \dlstyle{1114}	& \dlstyle{111}
\end{tabular}
\vspace{-6mm}
\end{table}

We compare our previous FHD and UHD scenarios, with adding one home-caching device per home, and a very optimistic reduction of both the peak percentage of OTT users ($s_\vod$) and total hours of OTT videos by 50\% for the FHD scenario, and 25\% for the UHD one. This difference could be explained by the fact that FHD videos being significantly smaller, many more videos could be cached in the same device. Yearly energy consumption results are summarized in Table~\ref{tab:dtt}.
In both scenarios, the additional power consumption induced by the home-caching devices cannot counter-balance the minor absolute gains. This result contradicts the conclusions of the LoCaT study which is based on misleading Wh/GB energy intensity indicators. Indeed, as already observed and as confirmed again in this table, reducing by 25\% or 50\% the OTT video traffic (both in volume and peak bitrate), yields to rather small absolute reduction of the global energy consumption. Surprisingly, a 25\% reduction on the bandwidth intensive UHD scenario leads to an energy reduction of 5.4\%, whereas a higher usage reduction of 50\% on the FHD scenario leads to a smaller relative energy reduction of 3\%. Yet another example of how relative indicators can be counter-intuitive.

\begin{table}[b]
\centering
\vspace{-5mm}
\caption{OTT versus \{DTT+ Home caching devices\} scenarios}
\label{tab:dtt}
\begin{tabular}{c|cc|cc}
 Scenario& \multicolumn{2}{c|}{OTT} & \multicolumn{2}{c}{OTT + DTT caching } \\
	     & $s_{\vod}$ & GWh & $s_{\vod}$ & GWh (network+home-cache) \\
\hline
FHD	& 20\% & \textbf{790}	& 10\% & \textbf{1271} (768+502) \\
UHD	& 20\% &\textbf{901}	& 15\% & \textbf{1362} (860+502) \\
\end{tabular}
\end{table}

Finally, these different results suggest that an ideal strategy for such a small caching device would be to insert it directly within the OLTs with a direct forward connection to the GPON cards. This way the gains of this cache on the rest of the network elements (10GE uplinks, edge, core, CDN, etc.) would be preserved, but the additional power consumption of a single caching device would be shared by thousands of users. Using a conservative number of 8000 subscribers per OLT, and a constant power of 30W per caching device would yield a negligible additional energy consumption of about $0.7$ GWh. Our model also reveals that filling this cache through DTT or over the network during low traffic periods will not make any practical difference regarding the energy consumption of the rest of the network. Note that this option is purely speculative since it is unclear whether this is even technically possible, and what would be the actual power of such a caching card.

\section{Discussions}
In this section, we discuss some choices made, limits and future work for our model. 

\paragraph{Conservative hypotheses}
We emphasize that our observations on the non-proportionality (even for the relative plot of Figure~\ref{fig:diff}) and on the sobriety vs efficiency conflict would be exacerbated through less conservative choices such as:
\begin{itemize}
\item Home-routers: we considered a very efficient ONU (2.5 W) but adding the associated home-router energy consumption would significantly flatten the absolute consumption variations.
\item Peak correlation: decorrelating the baseline and VoD streaming peaks would allow the baseline infrastructure to better absorb part of the VoD traffic. 
\item Homogeneous tree: a more realistic non-homogeneous distribution of the population would result in some nodes of the baseline scenario to be overdimensioned, hence leading to better handling of the additional VoD traffic.
\item Homogeneous technology: all our scenarios are based on the same building elements with a rather low granular capacity. Basing the baseline infrastructure on bleeding-edge technologies (e.g., WDM at 100 Gbps per chanel, 10 Gbps GPON) would again yield an overdimensioned baseline network, whereas allowing the most traffic-intensive scenarios to use different technologies would flatten the variations. This last option should, however, be accompanied with an increase of embodied impacts because of anticipated renewal.
\end{itemize}
We also assumed that the ONUs are always on. Enabling ONU/home-routers to be switched off when unneeded would result in a significant reduction of the absolute energy consumption. It is thus interesting to question how much a given usage is preventing such equipments to be switched off. In this regard, VoD streaming has a rather low impact with an average of 3.2h a day in our scenarios. For the baseline usage, a reasonable assumption would be to assume that they are switched off a few hours over nights and when the households are empty. An obvious worst-case usage is, however, smart-home equipment that require a permanent connection.



\paragraph{Complexity of the real world}
Just like previous bottom-up model, ours can only offer a simplified vision of the reality which is much more complex. For instance, network equipments exhibit a huge variability both in terms of capacity and efficiency. Equipments evolve with time with increased capacity and efficiency. With many actors deploying network equipment, peak bitrate demand is not the only driver for increase of the infrastructure, but economical competition and geopolitical strategies also play an important role leading to overdimensioning. As future work, it would thus be interesting to integrate all those aspects in such a model.


\paragraph{Restricted perimeters}
In this work, we have not included user devices, datacenters, nor content creation and encoding. On the datacenter side, maximal video resolution and quality is expected to have a significant effect on the computing (encoding) and storage resources. Those parameters are also expected to play an indirect but important role in accelerating the renewal of end-user equipements, for instance for larger 4K, HDR-enabled TVs, hence increasing the overall electricity consumption, but also manufacturing impacts.

\paragraph{Carbon footprint}
Converting energy consumption to carbon footprint requires knowing emission factors. These factors are country-dependent but also time-dependent as the energy mix depends on the time or season. Moreover averaged emission factors, even if made temporally varying, are not necessarily correlated to the \textit{consequential} effect of adding or removing a large body of electricity demand. Therefore, in this study, we have omitted this step on purpose.

\paragraph{Life cycle assessment}
Most studies are limited to the use phase of equipments, omitting other phases of their life cycle (material production, manufacturing, transport, installation, maintenance, end-of-life). The main reasons are the high level of uncertainty of estimating the emissions of these other phases. Some works do attempt to include embodied impacts~\cite{locat_2021,Ullrich2022} from an average ratio method that computes the scale of embodied emissions compared to the use phase.
In this paper, we have focused on the use phase only, but we acknowledge the strong importance of including the other phases, as well as accounting for other environmental impacts (e.g., water footprint, human toxicity, abiotic resource depletion, ...).
Properly allocating embodied emissions of shared equipment is as tricky as allocating static power consumption. In this regards, since our global approach bypasses the need for arbitrary allocation, we argue that the methodology proposed in this paper is well suited to be extended to estimate embodied impacts. Our model shall therefore be extended to enumerate all passive equipment that we have neglected so far (cables, shelters, buildings, racks, etc.), as well as installation and maintenance operations.

\section{Conclusion}
\label{sec:conclusion}

In this paper we have presented a novel methodology to assess the relationship between a given usage and network power consumption. Looking at the global energy consumption rather than attempting to arbitrarily allocate power between the different usages allowed us to avoid the classical pitfalls. Our results confirmed that classical efficiency indicators do not reflect the power consumption increase of more intensive Internet usage, and might even lead to misleading conclusions.
The bottom-up parametric network model we presented has the notable property of translating global average statistics to local smaller pools of inhabitants.
This theoretical network model is, however, necessarily imperfect and this paper discussed many future work opportunities such as variation of the density of population, variability of equipments, broader boundary, adding a broadband radio access network, and modeling other use-cases.
Another interesting future work would be to investigate how to extend our methodology to properly account for multiple use-cases whose peak demands are expected not to overlap.



\bibliographystyle{IEEEtran}
\bibliography{biblio.bib}
\end{document}